\documentclass[twocolumn]{jpsj2}

\def\runauthor{Y. NISHIWAKI and N. TODOROKI}

\title{Phase Diagram of Lattice-Spin System RbCoBr$_3$ 
}
\author{Yoichi NISHIWAKI\thanks{E-mail address: nishiwaki@lee.phys.titech.ac.jp}, Norikazu TODOROKI$^{1}$}

\inst{
Department of Physics (Cond.Mat.Phys.), Graduate School of Science and Engineering, 

Tokyo Institute of Technology, 2-12-1 Oh-okayama, Meguro-ku, Tokyo 152-8551 \\

$^{1}$ Institute of Physics, Kanagawa University,

3-27-1, Rokkaku-bashi, Kanagawa-ku, Yokohama, Kanagawa 221-8686
}

\recdate{\today}

\abst{We study the lattice-spin model of RbCoBr$_3$ which is proposed by Shirahata and Nakamura, by mean field approximation. This model is an Ising spin system on a distorted triangular lattice. There are two kinds of frustrated variables, that is, the lattice and spin. We obtain a phase diagram of which phase boundary is drawn continuously in a whole region. Intermediate phases that include a partial disordered state appear. The model has the first-order phase transitions in addition to the second-order phase transitions. We find a three-sublattice ferrimagnetic state in the phase diagram. The three-sublattice ferrimagnetic state does not appear when the lattice is not distorted.
}

\kword{RbCoBr$_3$, lattice-spin model, Ising spin system on distorted triangular lattice, frustration, mean field approximation}

\begin{document}

\maketitle

\section{Introduction}
\begin{figure}[t]
\begin{center}
\includegraphics[width=7cm]{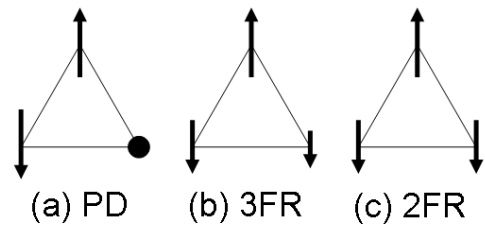}
\end{center}
\caption{Schematic sublattice magnetizations of (a) PD, (b) 3FR and (c) 2FR phases. Arrows indicate the direction and the magnitude of the sublattice magnetization. Black circle indicates that the sublattice magnetization is zero.}
\label{fig;fig1.eps}
\end{figure}

\begin{figure}[t]
\begin{center}
\includegraphics[width=7cm]{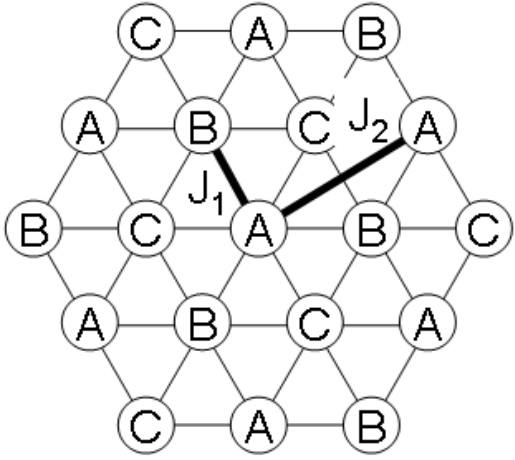}
\end{center}
\caption{Exchange interactions and sublattice of triangular lattice model. The NN interaction $J_1$ and the NNN interaction $J_2$ are shown. (A, B, C) denotes three different sublattices, as described in Appendix.}
\label{fig;fig2.eps}
\end{figure}

A large number of experimental and theoretical studies have been carried out for hexagonal ABX$_3$-type substances such as CsCoCl$_3$ and CsCoBr$_3$ because of the successive magnetic phase transitions due to the strong spin frustration.\cite{Achiwa,Mekata77,Mekata78,Yelon,Shiba,Todoroki} These substances consist of quasi-one-dimensional spin chains because the intrachain antiferromagnetic exchange interaction is much stronger than the interchain antiferromagnetic exchange interaction. Substances of this type are expressed by the stacked triangular antiferromagnetic Ising model. This model has three types of ordered phases: a partial disordered (PD) phase, a three-sublattice ferrimagnetic (3FR) phase and a two-sublattice ferrimagnetic (2FR) phase.\cite{Mekata78} These ordered phases are schematically shown in Figs. \ref{fig;fig1.eps}(a), \ref{fig;fig1.eps}(b) and \ref{fig;fig1.eps}(c). In the PD phase, the spins on one of the three sublattices are disordered and the spins on the other two sublattices are ordered antiferromagnetically. In the 3FR phase, the spins on the two sublattices are ordered in the same direction and the spins on the last sublattice are ordered in the opposite direction. The magnitudes of the sublattice magnetizations are mutually different. In the 2FR phase, the magnitudes of the parallel sublattice magnetizations are equal. Mekata described the magnetic phase transitions of these substances as a two-dimensional model with a nearest neighbor (NN) interaction and a next nearest neighbor (NNN) ferromagnetic interaction, as shown in Fig. \ref{fig;fig2.eps}, by mean field approximation (MFA). Recently, it has been found that the 3FR phase does not appear on these substances.\cite{Todoroki,Todoroki2002}

\begin{figure}[t]
\begin{center}
\includegraphics[width=7cm]{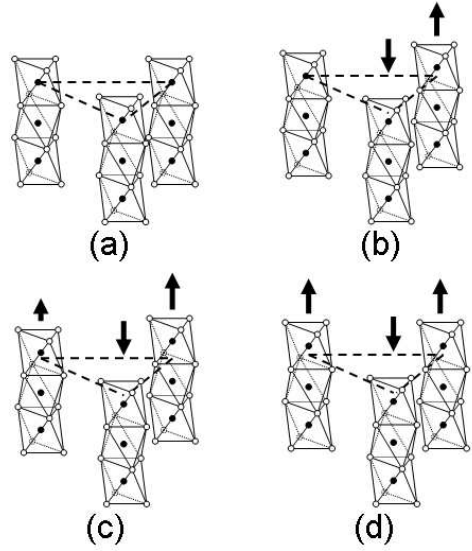}
\end{center}
\caption{Schematic manner of chain shifts of (a) lattice Para, (b) lattice PD, (c) lattice 3FR and (d) lattice 2FR phases. Black circles indicate the transition metal ions $B^{2+}$. Circles indicate the halogen ions $X^{-}$. The $X^{-}$ ions construct octahedra around $B^{2+}$ ions. Broken line triangles connect the positions of the $B^{2+}$ ions on the $c$-plane in the prototype structure. Here, the alkali ions $A^{+}$ are omitted for simplicity.}
\label{fig;fig3.eps}
\end{figure}

RbCoBr$_3$ is also a hexagonal ABX$_3$-type substance. RbCoBr$_3$ undergoes a structural phase transition  and a magnetic phase transition simultaneously at 37.0~K.\cite{Morishita,Yamanaka} The structural phase transition is characterized by an up or down shift of the -CoBr$_3$- chain structure along the $c$-axis. Recently, Shirahata and Nakamura have developed a model of RbCoBr$_3$ to explain this simultaneous structural-magnetic phase transition.\cite{Shirahata} They assumed that a shift of the Co$^{2+}$ ion along the $c$-axis could be treated as a pseudo Ising spin. They introduced a Hamiltonian consisting of two kinds of frustrated Ising spin variables, that is, a pseudo Ising spin of the lattice and a magnetic Ising spin. They considered that the exchange interaction changed with a relative position between two magnetic spins. They defined the ratio of the exchange interaction to an interaction of the pseudo Ising spin of the lattice by $\kappa$ and simulated this model  using a nonequilibrium relaxation method. They obtained a $T$-$\kappa$ phase diagram. The ordered phases of the lattice are schematically shown in Figs. \ref{fig;fig3.eps}(b), \ref{fig;fig3.eps}(c) and \ref{fig;fig3.eps}(d). We call these ordered phases of the lattice a "lattice PD" phase, a "lattice 3FR" phase and a "lattice 2FR" phase. In the lattice PD phase, the Co$^{2+}$ ions on one of the three sublattices are not shifted and the Co$^{2+}$ ions on the other two sublattices are shifted in the opposite direction along the $c$-axis. In the lattice 3FR phase, the Co$^{2+}$ ions on the two sublattices are shifted in the same direction and the Co$^{2+}$ ions on the last sublattice are shifted in the opposite direction. The magnitudes of the shifts of Co$^{2+}$ ions are mutually different. In the lattice 2FR phase, the magnitudes of the shifts of two Co$^{2+}$ ions are equal. They found that the simultaneous structural-magnetic phase transition occurred when the energy scale of the lattice part in the Hamiltonian coincided with the one of the spin part. However, there are two points which are not clear in their results. First, it seems that phase boundaries in their phase diagram cannot be drawn continuously near $\kappa = 0$. An examination of the intermediate phases may be insufficient. Second, they assumed that the phase transitions were of a second-order. They applied only the nonequilibrium relaxation method for the second-order transition. When the model has the first-order transitions, we have to use numerical methods which are applicable in the case of the first-order transition such as a mixed start nonequilibrium relaxation method.\cite{Ozeki}

Here, we study the lattice-spin model proposed by Shirahata and Nakamura, by MFA. Our aims are to clarify the above two points and to obtain a complete phase diagram.

\section{Model and Results}
We start with the following Hamiltonian proposed by Shirahata and Nakamura,
\begin{align}
{\cal H}={\cal H}_{\rm lattice}+{\cal H}_{\rm spin},
\label{eq:Hamiltonian}
\end{align}
where ${\cal H}_{\rm lattice}$ is an elastic part and ${\cal H}_{\rm spin}$ is a magnetic part. 

${\cal H}_{\rm lattice}$ is expected to take an expression of elastic energy with regard to the lattice variables,
\begin{align}
\begin{split}
{\cal H}_{\rm lattice}=& \sum_{i,j} J_0^{\rm L}(L_{ij}-L_{(i+1)j})^2\\
&+\sum_{i} \sum_{\left\langle jk \right\rangle}^{\rm NN}J_1^{\rm L}(L_{ij}-L_{ik})^2\\
&+\sum_{i} \sum_{\left\langle jl \right\rangle}^{\rm NNN}J_2^{\rm L}(L_{ij}-L_{il})^2.
\end{split}
\label{eq:lattice}
\end{align}
Here, $\sum_{\left\langle jk \right\rangle}^{\rm NN}$ and $\sum_{\left\langle jl \right\rangle}^{\rm NNN}$ run over NN pairs and NNN pairs on the $c$-plane, respectively. We assume for simplicity that the lattice variable $L_{ij}$ takes +1 or $-$1 depending on whether the Co$^{2+}$ ion shifts upward or downward along the $c$-axis from the prototype structure. This treatment is different from the previous study where $L_{ij}=\pm1$ and 0.\cite{Shirahata} However, this difference has little influence on the results qualitatively. The subscript $i$ denotes a site along the $c$-axis, and $j$ denotes a site on the $c$-plane. $J_0^{\rm L}$, $J_1^{\rm L}$ and $J_2^{\rm L}$ are spring constants of the NN pairs along the $c$-axis, the NN pairs on the $c$-plane and the NNN pairs on the $c$-plane, respectively. $J_0^{\rm L}$ is positive and   stronger than $J_1^{\rm L}$ and $J_2^{\rm L}$ because the chain structure is very hard. $J_1^{\rm L}$ is negative and $J_2^{\rm L}$ is positive.

${\cal H}_{\rm spin}$ is defined by the following expression with the exchange interaction dependent on the lattice variables,
\begin{align}
\begin{split}
{\cal H}_{\rm spin}=& -\sum_{i,j} (J_0^{\rm S}-\triangle J_0^{\rm S} |L_{ij}-L_{(i+1)j}|)S_{ij} S_{(i+1)j}\\
&-\sum_{i} \sum_{\left\langle jk \right\rangle}^{\rm NN}(J_1^{\rm S}-\triangle J_1^{\rm S} |L_{ij}-L_{ik}|)S_{ij} S_{ik}\\
&-\sum_{i} \sum_{\left\langle jl \right\rangle}^{\rm NNN}(J_2^{\rm S}-\triangle J_2^{\rm S} |L_{ij}-L_{il}|)S_{ij} S_{il}.
\end{split}
\end{align}
Here, we suppose that the Ising spin variable $S_{ij}$ takes +1 or $-$1. $J_0^{\rm S}$, $J_1^{\rm S}$ and $J_2^{\rm S}$ are the exchange interactions of the NN pairs along the $c$-axis, the NN pairs on the $c$-plane and the NNN pairs on the $c$-plane, respectively, in the prototype structure. We assume that the deformation of the lattice always decreases the magnitude of the exchange interaction. For simplicity, the decrease in the exchange interaction is supposed to be proportional to an absolute value of difference in the lattice variables: $|L_{ij}-L_{i^{\prime} j^{\prime}}|$, and coefficients are $\triangle J_{(0,1,2)}^{\rm S}$. These  parameters $\triangle J_{(0,1,2)}^{\rm S}$ control effective coupling between the lattice and the spin. $J_1^{\rm S}$ is negative and $J_2^{\rm S}$ is positive.\cite{Mekata77} The intrachain exchange interaction of RbCoBr$_3$ is antiferromagnetic. Here, we perform the unitary transformation $US_iU^{\dagger}=(-1)^iS_i$, which changes the sign of the intrachain interaction. Thus we consider that $J_0^{\rm S}$ and $\triangle J_0^{\rm S}$ are positive. $J_0^{\rm S}$ is stronger than $J_1^{\rm S}$ and $J_2^{\rm S}$.

\begin{figure}[t]
\begin{center}
\includegraphics[width=7cm]{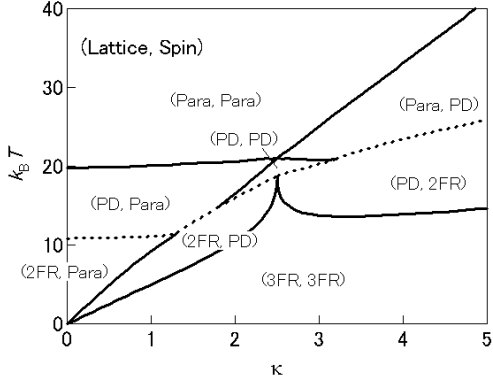}
\end{center}
\caption{$T$-$\kappa$ phase diagram. The broken and solid lines show that the phase transitions are of the first-order and of the second-order, respectively.}
\label{fig;fig4.eps}
\end{figure}

We apply to the above Hamiltonian a Scalapino-Imry-Pincus theory and solve it numerically. This theory explains the experimental results of CsCoCl$_3$ quantitatively.\cite{Shiba} It is expected that this theory is valid for the present model. The application of this method is shown in Appendix. We define a ratio of $J^{\rm S}$ to $J^{\rm L}$ by $\kappa=J^{\rm S}/J^{\rm L}$ and use
\begin{subequations}
\begin{align}
J_0^{\rm L}=5,~J_1^{\rm L}&=-1,~J_2^{\rm L}=0.1,\\
J_0^{\rm S}=5\kappa,~J_1^{\rm S}&=-\kappa,~J_2^{\rm S}=0.1\kappa,\\
\triangle J_{(0,1,2)}^{\rm S}&=0.2J_{(0,1,2)}^{\rm S}.
\end{align}
\end{subequations}
The obtained $T$-$\kappa$ phase diagram is shown in Fig. \ref{fig;fig4.eps}. When the lattice state is the lattice 2FR and the spin state is the PD, the phase is referred to as (lattice, spin)=(2FR, PD). We call the prototype structure and  paramagnetic state the lattice Para and magnetic Para, respectively. The broken and solid lines show that the phase transitions are of the first-order and of the second-order, respectively.

\section{Discussion}
We obtain the $T$-$\kappa$ phase diagram of the Shirahata-Nakamura's lattice-spin model by MFA. The previous study did not distinguish the 3FR state from the 2FR state, and called both the 3FR and 2FR states a Ferri state.\cite{Shirahata} This study shows that the (Ferri, Ferri) phase in the previous study is a (3FR, 3FR) phase. The other Ferri states in the previous study are 2FR states. The phase boundaries in the previous study intersected at one point. Then, the simultaneous structural-magnetic phase transition from (Para, Para) to (Ferri, Ferri) occurred. However, the phase boundaries in this study does not intersect at one point because of the presence of intermediate (PD, Para), (Para, PD) and (PD, PD) phases. These phases were not observed in the previous study. In the case of $\kappa=2.5$ given by
\begin{subequations}
\begin{align}
2J_0^{\rm L}&=J_0^{\rm S}-\triangle J_0^{\rm S},
\\
2J_1^{\rm L}&=J_1^{\rm S}-\triangle J_1^{\rm S},
\\
2J_2^{\rm L}&=J_2^{\rm S}-\triangle J_2^{\rm S},
\end{align}
\end{subequations}
a lattice part of the Hamiltonian coincides with a spin part. (See eqs.(\ref{HamiltonianL}), (\ref{HamiltonianS}), (\ref{HamiltonianJL}), (\ref{HamiltonianHL}), (\ref{HamiltonianJS}) and (\ref{HamiltonianHS}) in Appendix.) Then, successive  simultaneous structural-magnetic phase transitions (Para, Para) $\to $ (PD, PD) $\to $ (3FR, 3FR) occur. In the case of approximately $1.8<\kappa <2.5$, the phase transitions (Para, Para) $\to $ (PD, Para) $\to $ (PD, PD) $\to $ (2FR, PD) $\to $ (3FR, 3FR) occur. In the case of approximately $1.3<\kappa <1.8$, the phase transitions (Para, Para) $\to $ (PD, Para) $\to $ (2FR, PD) $\to $ (3FR, 3FR) occur. In the case of approximately $\kappa<1.3$, the phase transitions (Para, Para) $\to $ (PD, Para) $\to $ (2FR, Para) $\to $ (2FR, PD) $\to $ (3FR, 3FR) occur. The phase transitions from (PD, Para) to (2FR, PD) and from (2FR, PD) to (3FR, 3FR) are also simultaneous structural-magnetic phase transitions. The simultaneous structural-magnetic phase transition occurs in a wide $\kappa$ range, although it occurs only at one point in the previous study. That is, the simultaneous structural-magnetic phase transition tends to occur when the interaction of the lattice and that of the spin are comparable. In the case of $\kappa >2.5$, the behaviors of the lattice and the spin are reversed.

In the case of the only lattice system, $\kappa=0$, the successive structural phase transitions lattice Para $\to $ PD $\to $ 2FR occur. It seemed that the phase boundaries in the previous study could not be drawn continuously near $\kappa = 0$. This is because the (PD, Para) phase did not appear in the previous study. The (PD, Para), (Para, PD) and (PD, PD) phases must appear to draw continuous phase boundaries in the whole region of the phase diagram. When the lattice is not distorted, the fluctuation narrows the temperature range of the magnetic PD phase which is obtained by MFA, but the magnetic PD phase does not disappear. Therefore, it seems reasonable to assume that the phases which include the lattice or magnetic PD state do not disappear, even if we consider the fluctuation. Because the initial configurations (Ferri, PD), (Ferri, Para), (PD, PD) and (Para, Para) in the previous study seem insufficient, the (PD, Para), (Para, PD) and (PD, PD) phases might disappear. It is necessary that we use additional initial configurations, such as the (PD, Para) and (Para, PD) phases, for the nonequilibrium relaxation method.

These results show the first-order phase transitions in addition to the second-order phase transitions. The phase transitions   from lattice or magnetic PD to 2FR are of the first-order. When the lattice is not distorted, phase transitions from the magnetic Para to PD and from the PD to 2FR are of the second-order and of the first-order, respectively.\cite{Todoroki} These  results are consistent. In the previous study, this model was examined using only the nonequilibrium relaxation method for the second-order phase transition. It is necessary that we use methods which are also applicable to the first-order phase transition to discuss this model.

The lowest-temperature phase is the (3FR, 3FR) phase. The magnetic 3FR state does not appear when the lattice is not distorted.\cite{Todoroki} Lattice and spin degrees of freedom and nonequivalence interactions induce the appearance of (3FR, 3FR) phase. 

\section*{Acknowledgment}

The authors would like to thank T. Nakamura, S. Miyashita, K. Iio and H. Tanaka for useful discussions. This work was supported by a 21st Century COE Program at Tokyo Tech "Nanometer-Scale Quantum Physics" by the Ministry of Education, Culture, Sports, Science and Technology.

\appendix
\section{Mean Field Approximation}
We apply to eq. (\ref{eq:Hamiltonian}) the Scalapino-Imry-Pincus theory, that is, we use the following MFA,
\begin{subequations}
\begin{align}
L_{i}^{\rm A} L_{i}^{\rm B} \sim & \ l^{\rm B} L_{i}^{\rm A} +l^{\rm A} L_{i}^{\rm B}-l^{\rm A}l^{\rm B},\\
S_{i}^{\rm A} S_{i}^{\rm B} \sim & \ m^{\rm B} S_{i}^{\rm A} +m^{\rm A} S_{i}^{\rm B}-m^{\rm A}m^{\rm B},\\
\begin{split}
L_{i}^{\rm A} L_{i}^{\rm B} S_{i}^{\rm A} S_{i}^{\rm B} \sim & \ l^{\rm B} m^{\rm A} m^{\rm B} L_{i}^{\rm A} +l^{\rm A}m^{\rm A} m^{\rm B} L_{i}^{\rm B}\\
&+l^{\rm A}l^{\rm B} m^{\rm B} S_{i}^{\rm A} +l^{\rm A}l^{\rm B} m^{\rm A} S_{i}^{\rm B}\\
&-3l^{\rm A}l^{\rm B}m^{\rm A}m^{\rm B},
\end{split}\\
\begin{split}
L_{i}^{\rm A} L_{i+1}^{\rm A} S_{i}^{\rm A} S_{i+1}^{\rm A} \sim & \ \eta^{\rm A}_{\rm S} L_{i}^{\rm A} L_{i+1}^{\rm A}+\eta^{\rm A}_{\rm L} S_{i}^{\rm A} S_{i+1}^{\rm A}\\
&-\eta^{\rm A}_{\rm L} \eta^{\rm A}_{\rm S},
\end{split}
\end{align}
\end{subequations}
where, (A, B and C) denote three different sublattices, as shown in Fig. \ref{fig;fig2.eps}. We introduce sublattice moments as
\begin{subequations}
\begin{align}
l^{\lambda}&=\frac{1}{N_{\rm 1D}}\sum_{i} \left\langle L_{i}^{\lambda} \right\rangle,\\
m^{\lambda}&=\frac{1}{N_{\rm 1D}}\sum_{i} \left\langle S_{i}^{\lambda} \right\rangle,
\end{align}
\end{subequations}
and energy along the $c$-axis as
\begin{subequations}
\begin{align}
\eta_{\rm L}^{\lambda}&=\frac{1}{N_{\rm 1D}}\sum_{i} \left\langle L_{i}^{\lambda} L_{i+1}^{\lambda} \right\rangle,\\
\eta_{\rm S}^{\lambda}&=\frac{1}{N_{\rm 1D}}\sum_{i} \left\langle S_{i}^{\lambda} S_{i+1}^{\lambda} \right\rangle,
\end{align}
\end{subequations}
where, $N_{\rm 1D}$ is the number of spins in the chains, and $\lambda$ denotes A, B and C. The Hamiltonian for the A sublattice is written in the above MFA and can be divided into the lattice part and the spin part as,
\begin{subequations}
\begin{align}
{\cal H}_{\rm MFA}^{L^{\rm A}}&=J^{\rm L^{\rm A}} \sum_{i} L_{i}^{\rm A} L_{i+1}^{\rm A}+ H^{\rm L^{\rm A}} \sum_{i} L_{i}^{\rm A},\label{HamiltonianL}
\\
{\cal H}_{\rm MFA}^{S^{\rm A}}&=J^{\rm S^{\rm A}} \sum_{i} S_{i}^{\rm A} S_{i+1}^{\rm A}+ H^{\rm S^{\rm A}} \sum_{i} S_{i}^{\rm A},
\label{HamiltonianS}
\end{align}
\end{subequations}
where,
\begin{subequations}
\begin{align}
J^{\rm L^{\rm A}}=&-(2J_0^{\rm L}+\triangle J_0^{\rm S} \eta^{\rm A}_{\rm S} ),\label{HamiltonianJL}\\
\begin{split}
H^{\rm L^{\rm A}}=&-\{6J_1^{\rm L} ( l^{\rm B} + l^{\rm C} )+12 J_2^{\rm L} l^{\rm A}\\
&+3\triangle J_1^{\rm S} ( l^{\rm B} m^{\rm B} + l^{\rm C} m^{\rm C}) m^{\rm A}\\
&+6\triangle J_2^{\rm S} (m^{\rm A})^2 l^{\rm A} \},\label{HamiltonianHL}
\end{split}\\
J^{\rm S^{\rm A}}=&-(J_0^{\rm S}-\triangle J_0^{\rm S} +\triangle J_0^{\rm S} \eta^{\rm A}_{\rm L}),\label{HamiltonianJS}\\
\begin{split}
H^{\rm S^{\rm A}}=&-\{3(J_1^{\rm S}-\triangle J_1^{\rm S} )( m^{\rm B}+ m^{\rm C})\\
&+6(J_2^{\rm S}-\triangle J_2^{\rm S} ) m^{\rm A}\\
&+3\triangle J_1^{\rm S}( l^{\rm B} m^{\rm B} +
l^{\rm C} m^{\rm C} ) l^{\rm A}\\
&+6\triangle J_2^{\rm S} (l^{\rm A})^2 m^{\rm A} \}.\label{HamiltonianHS}
\end{split}
\end{align}
\end{subequations}
Equations (\ref{HamiltonianL}) and (\ref{HamiltonianS}) can be solved by the well-known exact solution of a one-dimensional Ising model. We obtain the following self-consistent equations,
\begin{subequations}
\begin{align}
l^{\rm A} &= \frac{\sinh (\beta H^{\rm L^{\rm A}})}{\sqrt{\exp (-4 \beta J^{\rm L^{\rm A}}) +\sinh^2 (\beta H^{\rm L^{\rm A}}})},
\label{eq;self1L}\\
m^{\rm A} &= \frac{\sinh (\beta H^{\rm S^{\rm A}})}{\sqrt{\exp (-4 \beta J^{\rm S^{\rm A}}) +\sinh^2 (\beta H^{\rm S^{\rm A}}})},
\label{eq;self1S}
\end{align}
\end{subequations}
and
\begin{subequations}
\begin{align}
\begin{split}
\eta^{\rm A}_{\rm L}=&\frac{1}{\tanh (2\beta J^{\rm L^{\rm A}})}\\
&-\frac{\cosh (\beta H^{\rm L^{\rm A}})}{\sinh (2\beta J^{\rm L^{\rm A}})\sqrt{\exp (4 \beta J^{\rm L^{\rm A}}) \sinh^2 (\beta H^{\rm L^{\rm A}})+1}},
\end{split}
\\
\begin{split}
\eta^{\rm A}_{\rm S}=&\frac{1}{\tanh (2\beta J^{\rm S^{\rm A}})}\\ 
&-\frac{\cosh \beta H^{\rm S^{\rm A}}}{\sinh (2\beta J^{\rm S^{\rm A}})\sqrt{\exp (4 \beta J^{\rm S^{\rm A}}) \sinh^2 (\beta H^{\rm S^{\rm A}})+1}},
\end{split}
\end{align}
\end{subequations}
where, $\beta=k_{\rm B}T$.

The partition functions of the lattice chain and spin chain are
\begin{subequations}
\begin{align}
\begin{split}
Z^{\rm L^{\rm A}}=&\exp(\beta J^{\rm L^{\rm A}}) \cosh (\beta H^{\rm L^{\rm A}})\\
&+\sqrt{\exp (2 \beta J^{\rm L^{\rm A}}) \sinh^2 (\beta H^{\rm L^{\rm A}})+\exp(-2\beta J^{\rm L^{\rm A}})},
\end{split}
\\
\begin{split}
Z^{\rm S^{\rm A}}=&\exp(\beta J^{\rm S^{\rm A}}) \cosh (\beta H^{\rm S^{\rm A}})\\
&+\sqrt{\exp (2 \beta J^{\rm S^{\rm A}}) \sinh^2 (\beta H^{\rm S^{\rm A}})+\exp(-2\beta J^{\rm S^{\rm A}})},
\end{split}
\end{align}
\end{subequations}
respectively. The free energy $F$ is given by
\begin{align}
\begin{split}
F=-N \beta \ln (Z^{\rm L^{\rm A}} Z^{\rm L^{\rm B}} Z^{\rm L^{\rm C}} Z^{\rm S^{\rm A}} Z^{\rm S^{\rm B}} Z^{\rm S^{\rm C}})+E_{\rm int},
\end{split}
\end{align}
where, $N$ is the numbers of three chains and
\begin{align}
\begin{split}
E_{\rm int}=&\frac{N}{2} \{ \triangle J_0^{\rm S} (\eta^{\rm A}_{\rm L} \eta^{\rm A}_{\rm S}
+ \eta^{\rm B}_{\rm L} \eta^{\rm B}_{\rm S}
+ \eta^{\rm C}_{\rm L} \eta^{\rm C}_{\rm S})\\
&+12 J_1^{\rm L} (l^{\rm A} l^{\rm B}+ l^{\rm B} l^{\rm C}+ l^{\rm C} l^{\rm A})\\
&+12J_2^{\rm L} \left[(l^{\rm A})^2 + (l^{\rm B})^2 +(l^{\rm C})^2 \right]\\
&+6(J_1^{\rm S}-\triangle J_1^{\rm S} ) (m^{\rm A} m^{\rm B}+ m^{\rm B} m^{\rm C}+m^{\rm C} m^{\rm A})\\
&+6(J_2^{\rm S}-\triangle J_2^{\rm S} ) \left[(m^{\rm A})^2 + (m^{\rm B})^2 + (m^{\rm C})^2 \right]\\
&+18\triangle J_1^{\rm S} (l^{\rm A} m^{\rm A} l^{\rm B} m^{\rm B}+ l^{\rm B} m^{\rm B} l^{\rm C} m^{\rm C}+l^{\rm C} m^{\rm C} l^{\rm A} m^{\rm A} )\\
&+18\triangle J_2^{\rm S} \left[(l^{\rm A})^2 (m^{\rm A})^2 + (l^{\rm B})^2 (m^{\rm B})^2 +(l^{\rm C})^2 (m^{\rm C})^2 \right]\}.
\end{split}
\end{align}
The $E_{\rm int}$ is necessary to correct the dual computation of the interaction.

\end{document}